\documentclass[aps,pra,twocolumn,showpacs,groupedaddress]{revtex4-1}
\usepackage{graphicx}
\usepackage{pstricks}     
\usepackage{amssymb}  
\usepackage{bm}

\begin{document}

\title{Parametric attosecond pulse amplification far from the ionization threshold from high order harmonic generation in He$^+$}

\author{C. Serrat}
\affiliation{Department of Physics, Polytechnic University of Catalonia, Colom 11, 08222 Terrassa (Barcelona), Spain}
\email{carles.serrat-jurado@upc.edu}
\author{J. Seres}
\author{E. Seres}
\affiliation{Institute of Atomic and Subatomic Physics, Vienna University of Technology, Stadionalle 2, 1020 Vienna, Austria}
\author{T-H. Dinh} 
\author{N. Hasegawa} 
\author{M. Nishikino}
\affiliation{Kansai Photon Science Institute, National Institutes for Quantum and Radiological Science and Technology (QST), Kizugawa, Kyoto 619-0215, Japan}
\author{S. Namba}
\affiliation{Graduate School of Engineering, Hiroshima University, 1-4-1 Kagamiyama, Higashi-Hiroshima, Hiroshima 739-8527, Japan}

 \date{\today}

\begin{abstract}
Parametric amplification of attosecond coherent pulses around 100 eV at the single-atom level 
is demonstrated for the first time by using the 3D time-dependent Schr{\"o}dinger equation
in high-harmonic generation processes from excited states of He$^+$. 
We present the attosecond dynamics of the amplification process 
far from the ionization threshold and resolve the physics behind it.
The amplification of a particular central photon energy requires the seed XUV pulses 
to be perfectly synchronized in time with the driving laser field for stimulated recombination 
to the He$^+$ ground state and is only produced in a few specific laser cycles in 
agreement with the experimental measurements. Our simulations show that 
the amplified photon energy region can be controlled
by varying the peak intensity of the laser field. Our results pave the way to the realization of 
compact attosecond pulse intense XUV lasers with broad applications.
\end{abstract}

\maketitle

\section{Introduction}

High harmonic generation (HHG) yield enhancement is needed for the design of compact 
XUV and x-ray lasers fitting in hospitals and university laboratories and therefore for 
HHG coherent radiation to be useful in a large number of applications \cite{Agostini1,Krausz1}. 
In particular, the amplification of attosecond HHG pulses at high photon energies 
in gases has been extensively investigated both for its fundamental interest and also for 
several applications, such as the use of parametric amplified HHG pulses as seed in full coherent plasma x-ray lasers 
\cite{Seres1,Seres2,Serrat1,Seres3,Seres4,Serrat2,Serrat3,Serrat4,Serrat5,Serrat6}. 
In the present study we investigate the amplification of attosecond pulses with specific central photon 
energies that are far from both the ionization energy of He (24.6 eV) and He$^+$ (54.4 eV). The 
results and the physical processes that we describe therefore essentially differ from other 
research on HHG yield and cutoff enhancement \cite{Ishikawa,Biegert,Brizuela,Kroh} or pulse
amplification near the ionization threshold \cite {Ivanov}.

\begin{figure*}[ht]
\begin{center}
\includegraphics[scale=0.17,clip=true,angle=0]{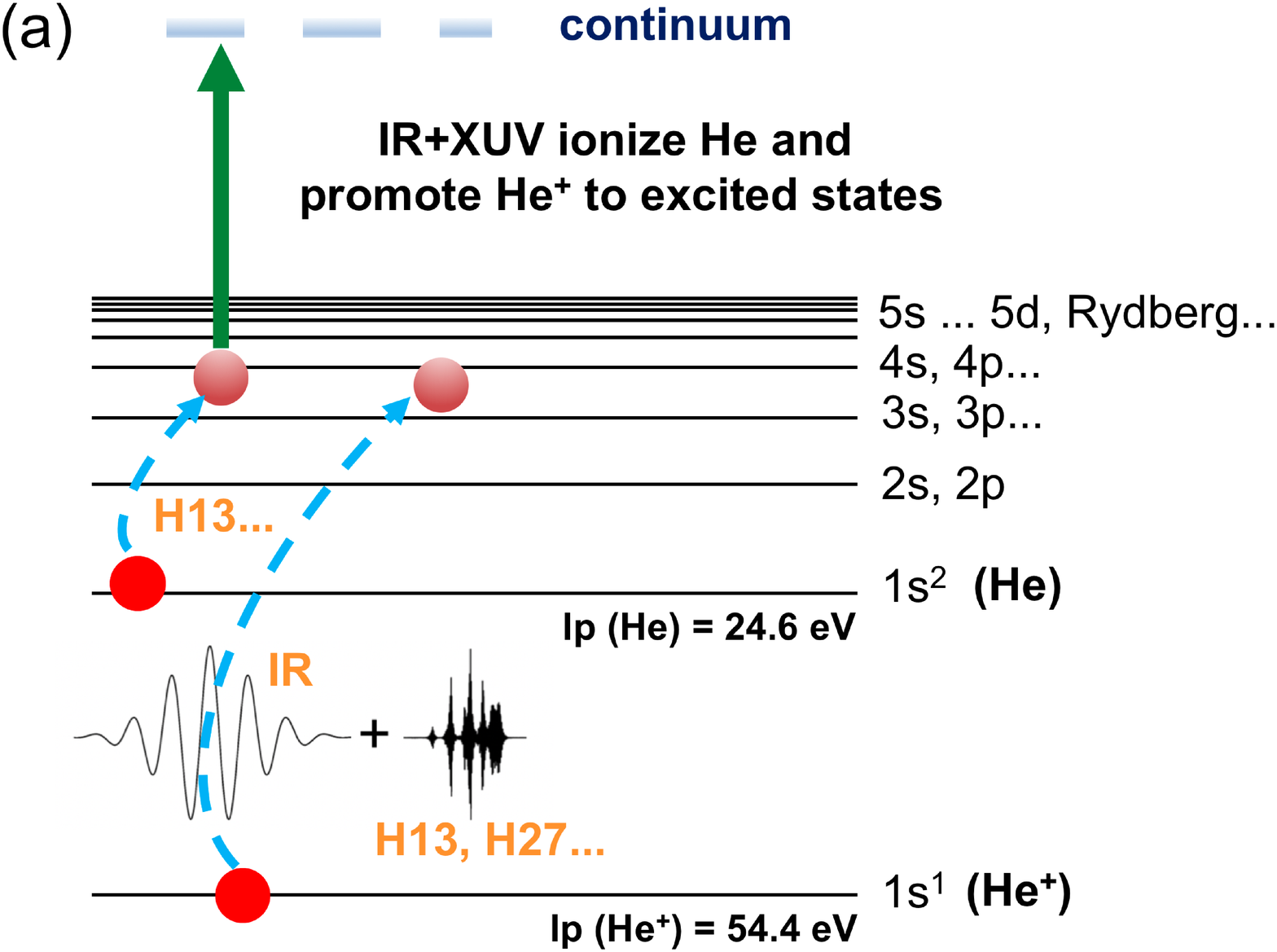}
\includegraphics[scale=0.17,clip=true,angle=0]{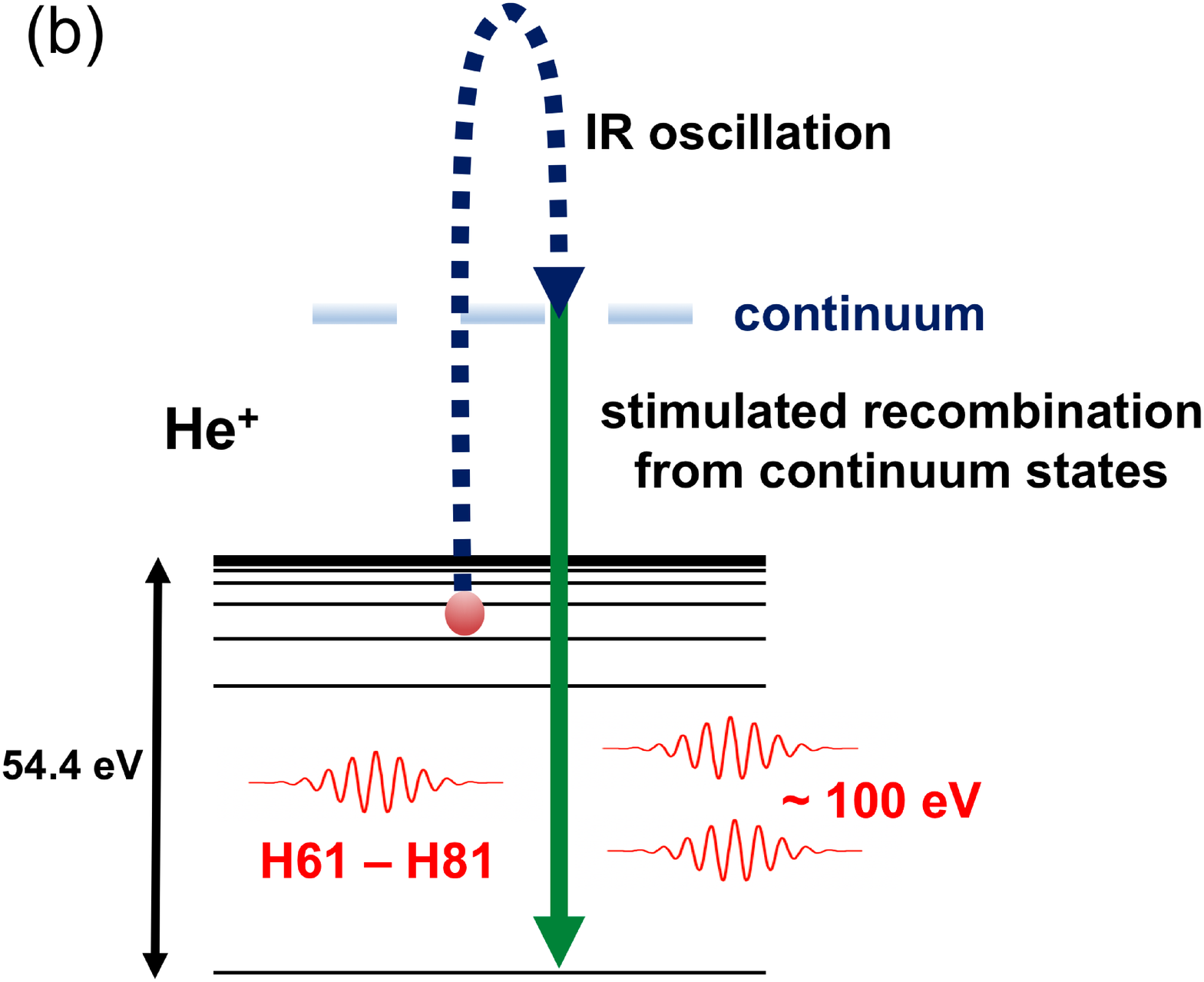}
\end{center}
\caption{(a) The combination of the IR and HHG XUV pulses produce He ions and promote 
He$^+$ to excited states. (b) He$^+$ ions initially in excited states oscillate in the intense IR laser field 
and the recombination from continuum states to the ground state of He$^+$ is stimulated by
XUV pulses optimally synchronized with the IR field.}
\label{fig1}
\end{figure*}
 
Attosecond XUV pulses of high photon energies
have been successfully amplified in experiments using He gas, and the fundamental physics behind the 
involved scattering mechanisms have been examined both experimentally and theoretically 
\cite{Seres1,Seres2,Seres3,Seres4,Serrat3,Serrat6}.
Parametric processes involving stimulated recombination in HHG have been proposed
as the dominant mechanism behind the observed XUV pulse amplification, 
although a detailed description at the single-atom level using a first principles theory allowing the 
understanding of the microscopic physical mechanisms had not been reported so far.
In this paper we examine the simultaneous excitation of He$^+$ ions 
with an intense IR pulse and a weak XUV pulse, by numerically 
solving the spin-free single-active electron (SAE) \cite{Muller1,Wiehle1} 3D time-dependent 
Schr{\"o}dinger equation using the implementation described in \cite{Serrat6,Patchkovskii1} with a hydrogen-like potential. 
Our simulations are compared with experimental measurements performed in a double-jet HHG experiment in He.
As it will be further detailed below, in the first jet the XUV seed is generated and is carefully synchronized with the 
intense IR pulse to interact afterwards with the He gas in the second jet, producing XUV amplification in a 
particular region of the HHG spectrum. At the considered peak intensities, the fundamental IR laser pulse 
does not substantially ionize the He medium. However, He ions can be easily produced in the second jet by the 
combination of the IR and XUV pulses, for instance through single-photon ionization by the H27 harmonics 
of the IR (800 nm) pulse or by assisting transitions to intermediate levels \cite{Ishikawa}.
In Fig. \ref{fig1} a schematic draw of the processes that are involved in the 
amplification in He is shown. Once the He atoms are ionized by the combination of the IR and 
HHG pulses, He$^+$ ions can be promoted to excited states and ionized by assisting transitions 
to intermediate levels [see Fig. \ref{fig1} (a)]. 
Extensive previous studies have shown that stimulated recombination to the ground state of He in HHG by energies
close to 100 eV is highly unlikely. Therefore, in the present study we consider
the He$^+$ ions initially in excited states and explain how an attosecond XUV pulse with a particular photon energy
near 100 eV can be parametrically amplified by stimulated recombination from continuum states to the ground state
of the He$^+$ ion, as schematically shown in  Fig. \ref{fig1} (b).

\begin{figure*}[ht]
 \centering
\begin{minipage}[t]{1.0\textwidth}
  \centering
  \includegraphics[width=0.32\textwidth]{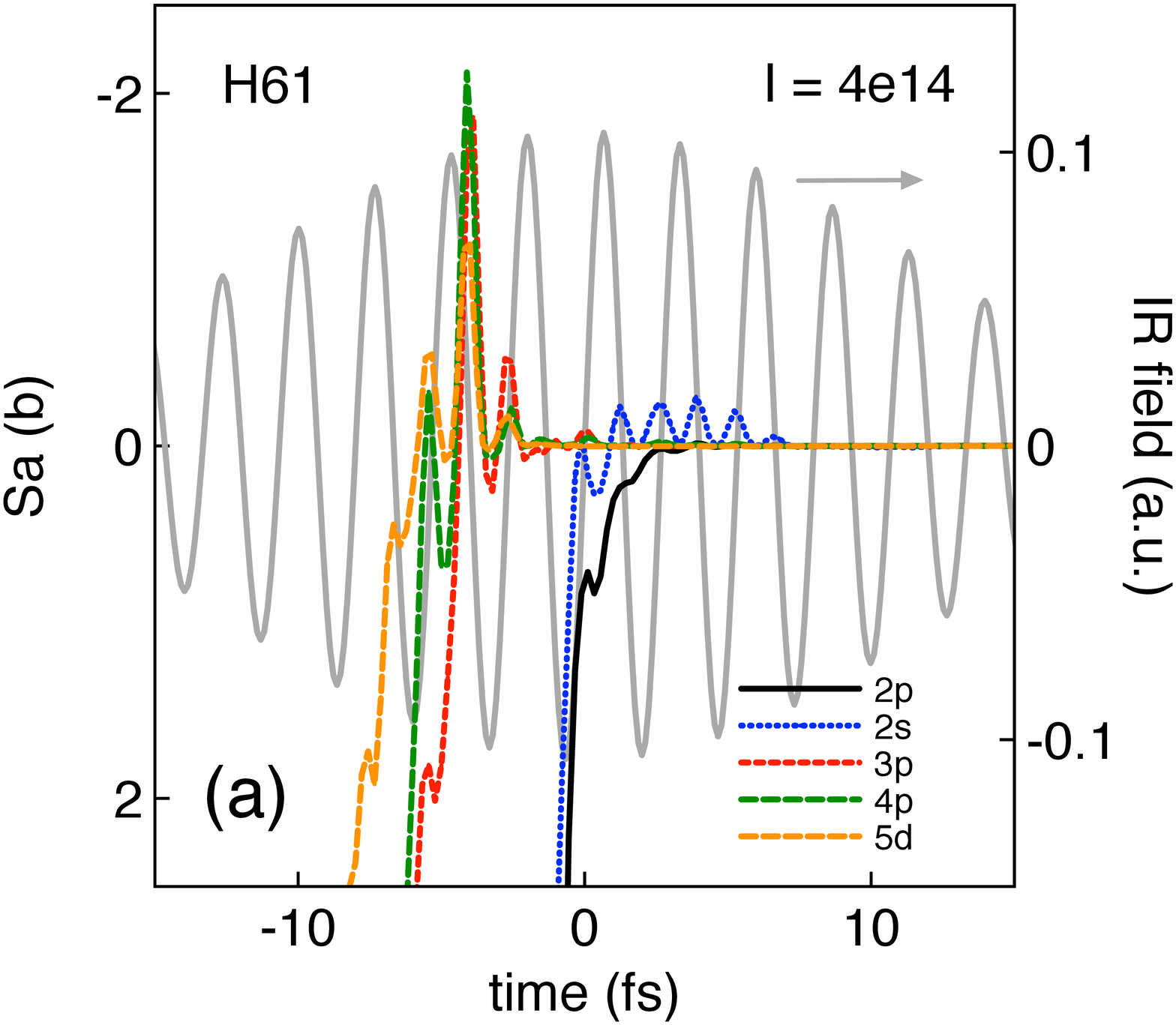} 
    \includegraphics[width=0.32\textwidth]{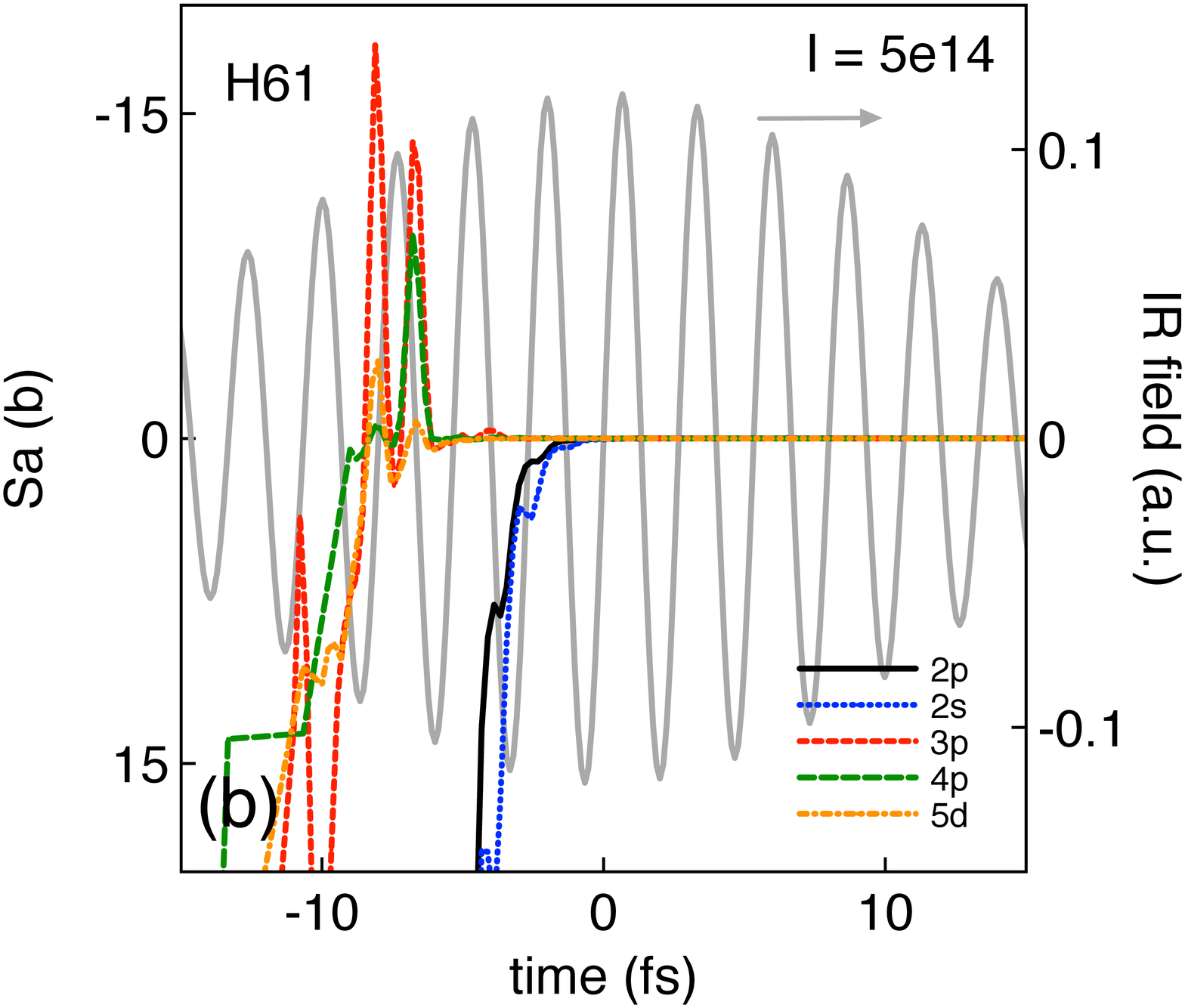}  
      \includegraphics[width=0.32\textwidth]{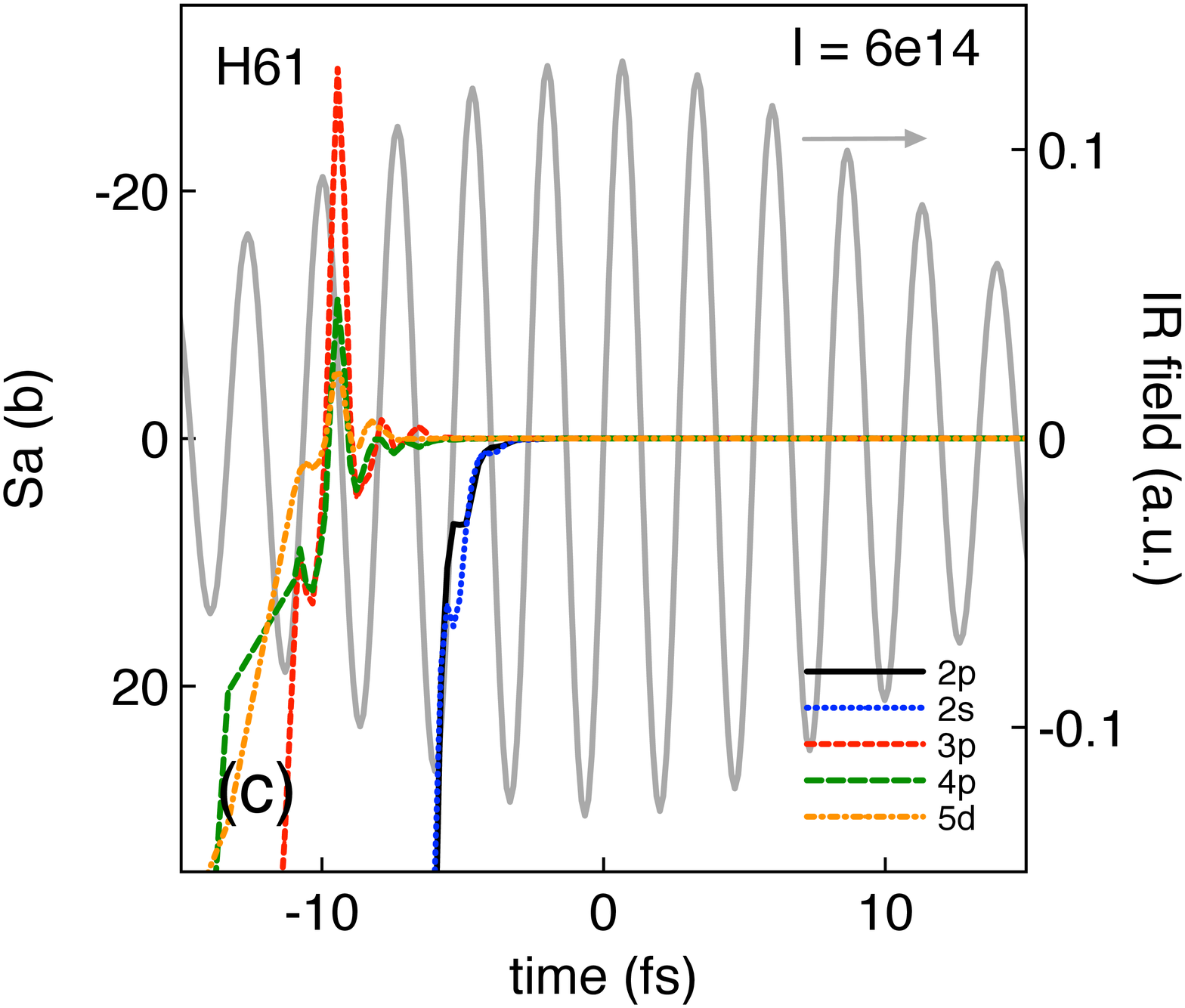}
\end{minipage}\qquad
\begin{minipage}[t]{1.0\textwidth}
  \centering
  \includegraphics[width=0.32\textwidth]{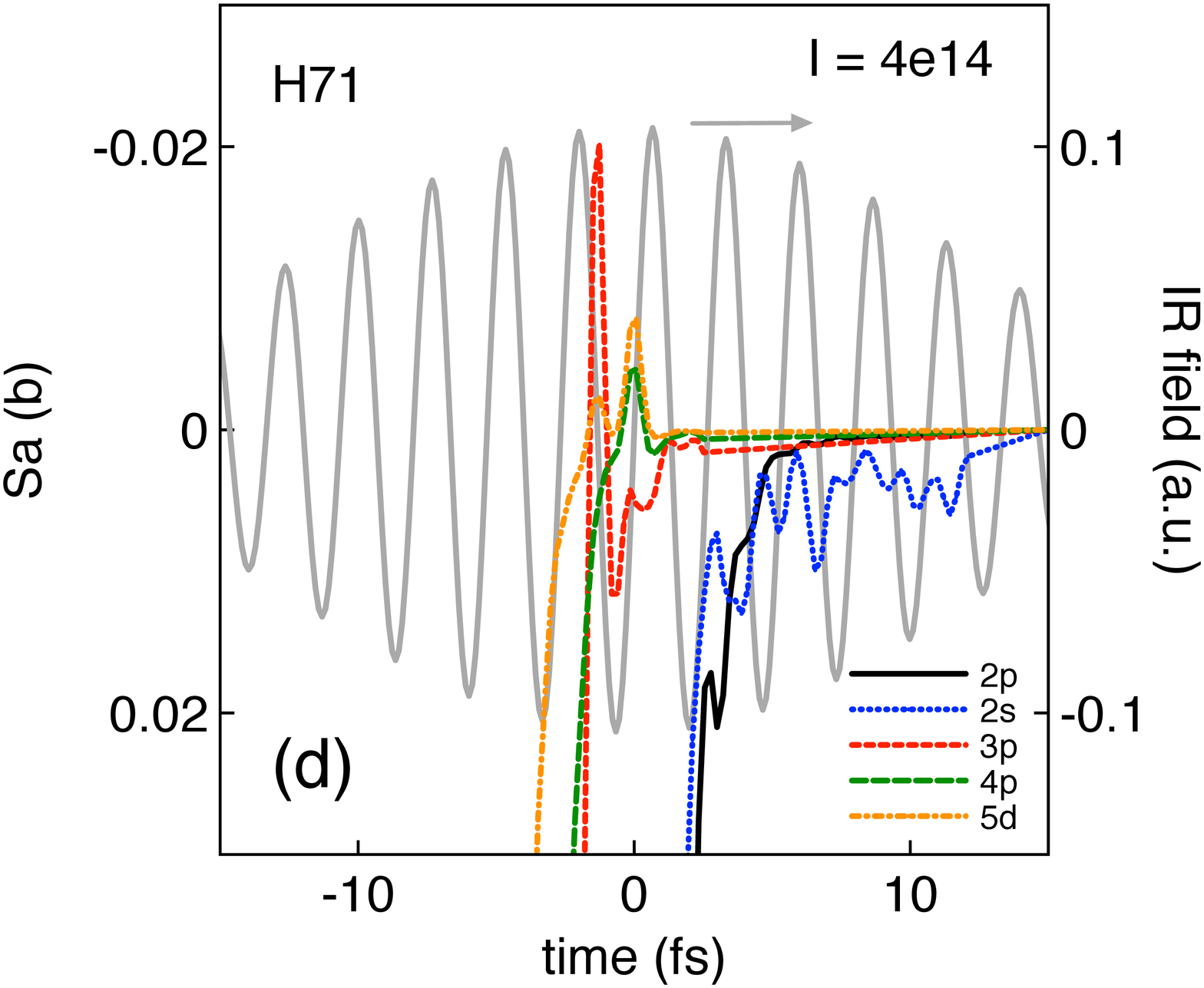} 
    \includegraphics[width=0.32\textwidth]{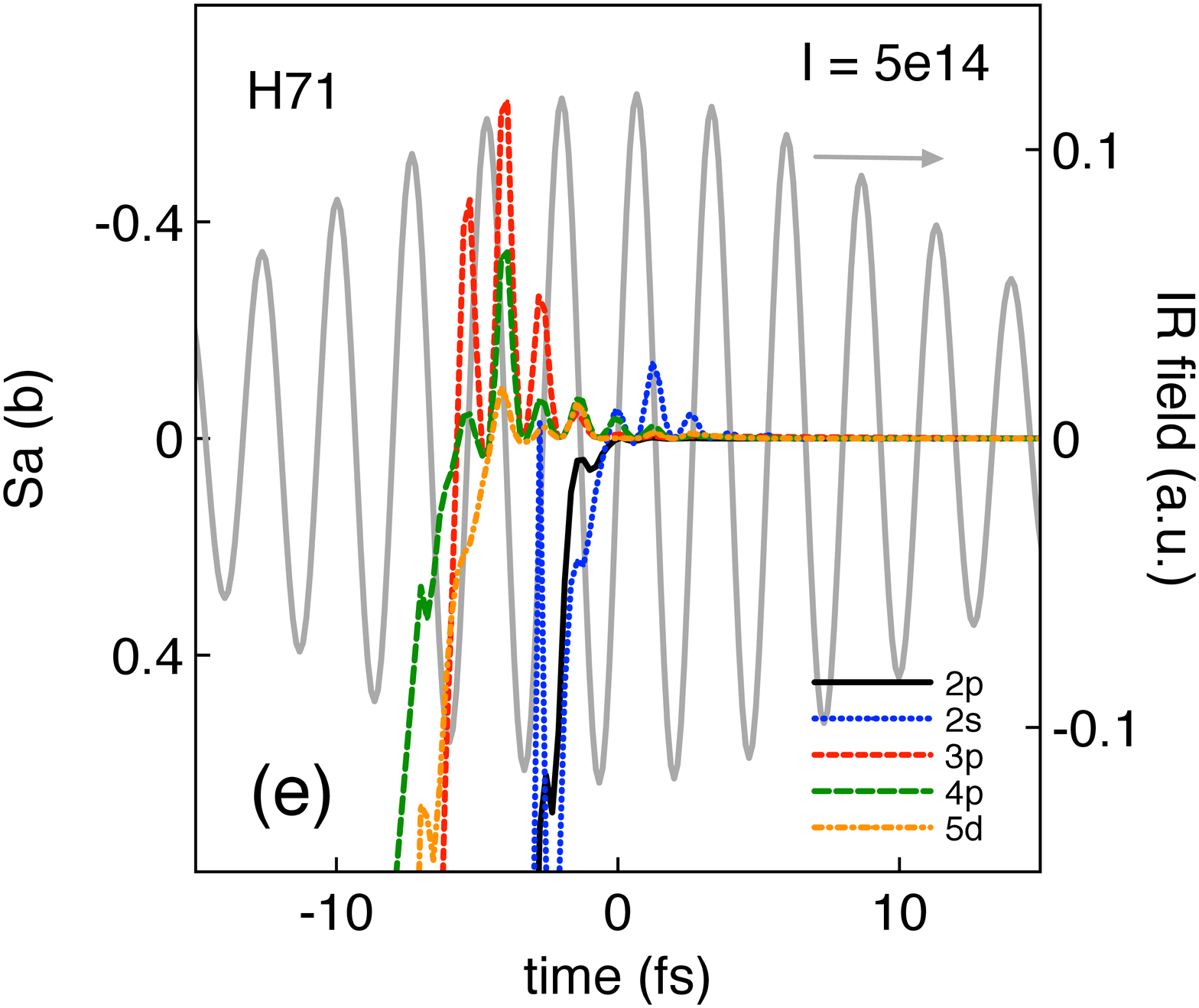}  
      \includegraphics[width=0.32\textwidth]{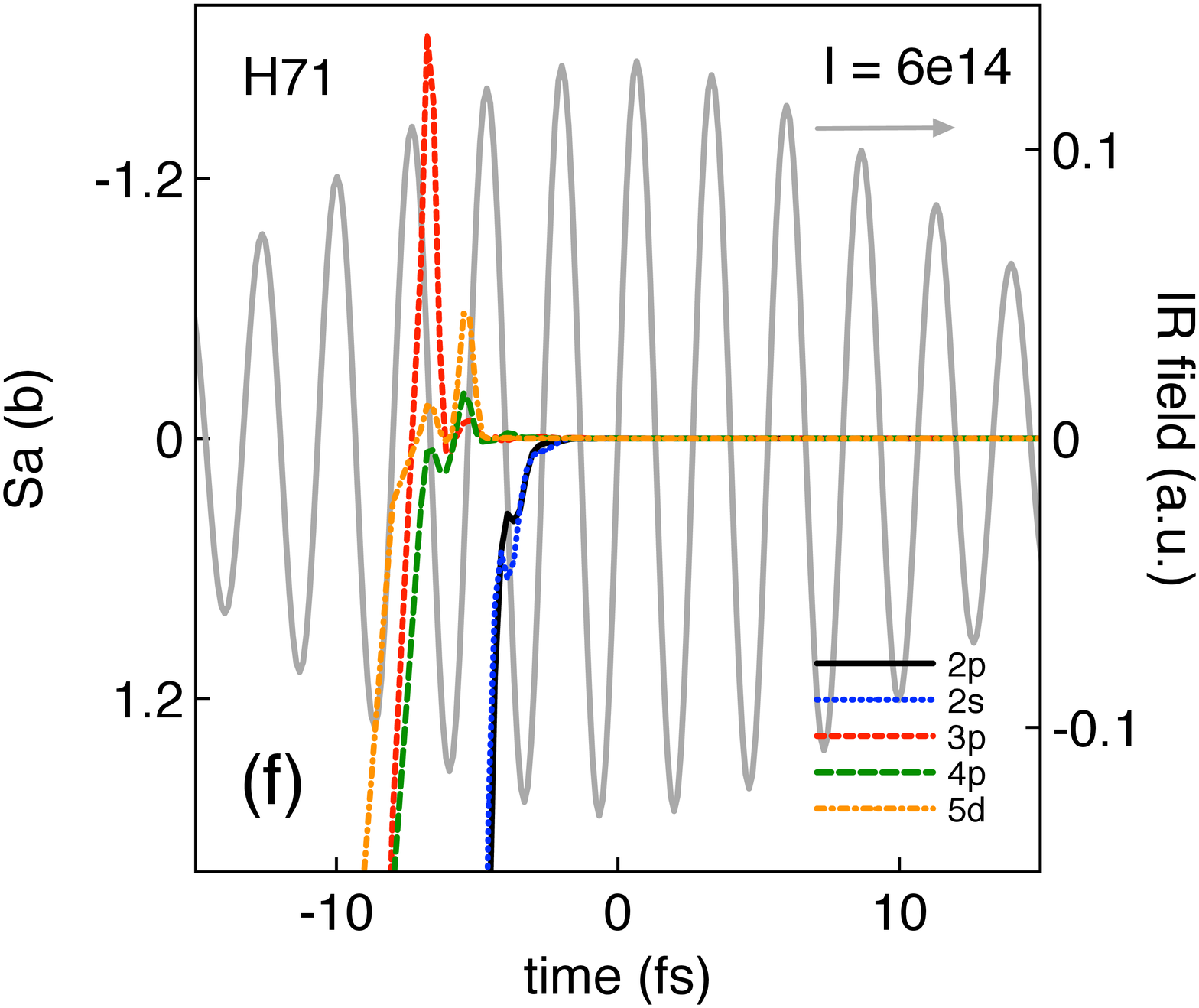}
\end{minipage}\qquad
\begin{minipage}[t]{1.0\textwidth}
  \centering
  \includegraphics[width=0.32\textwidth]{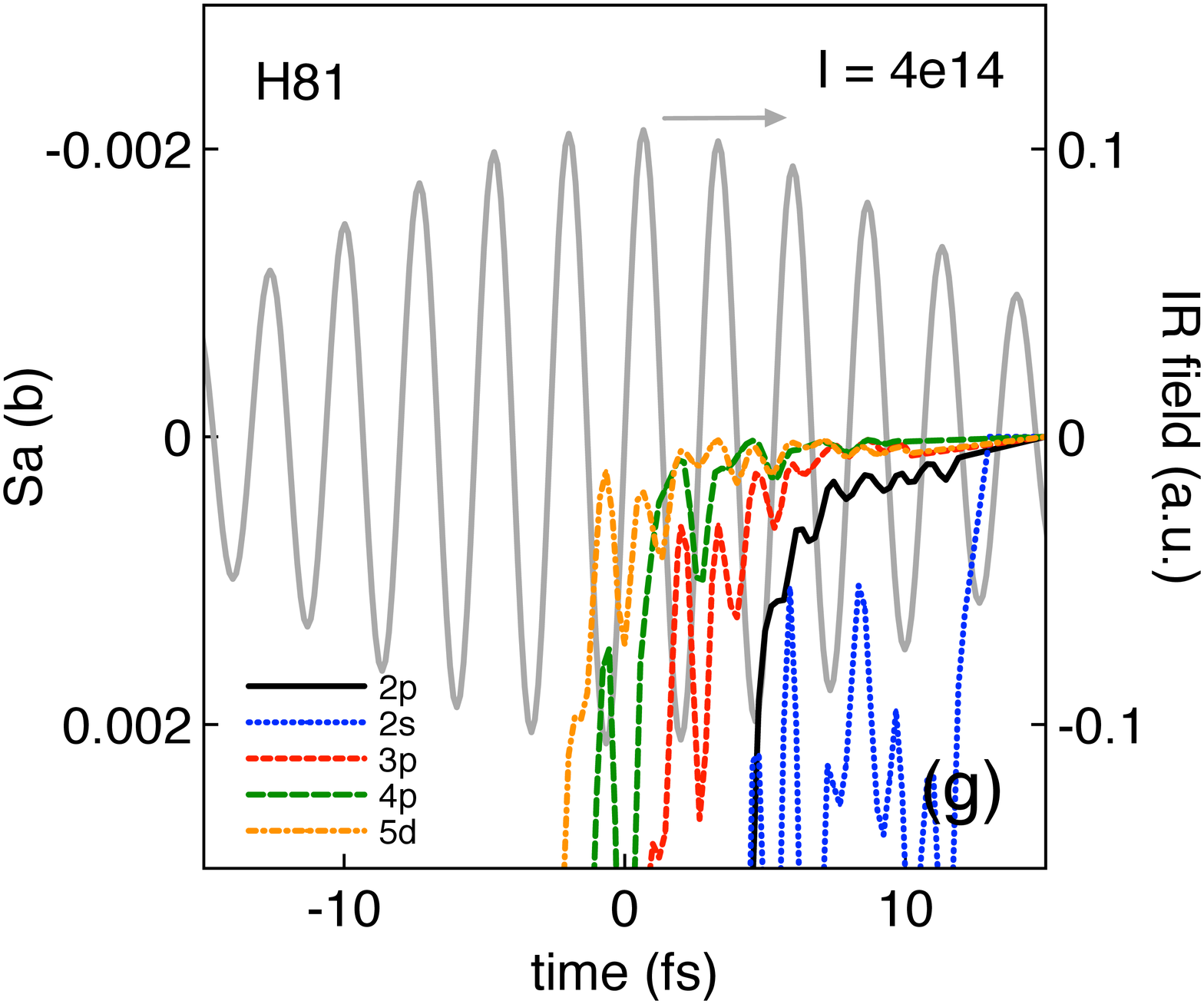} 
    \includegraphics[width=0.32\textwidth]{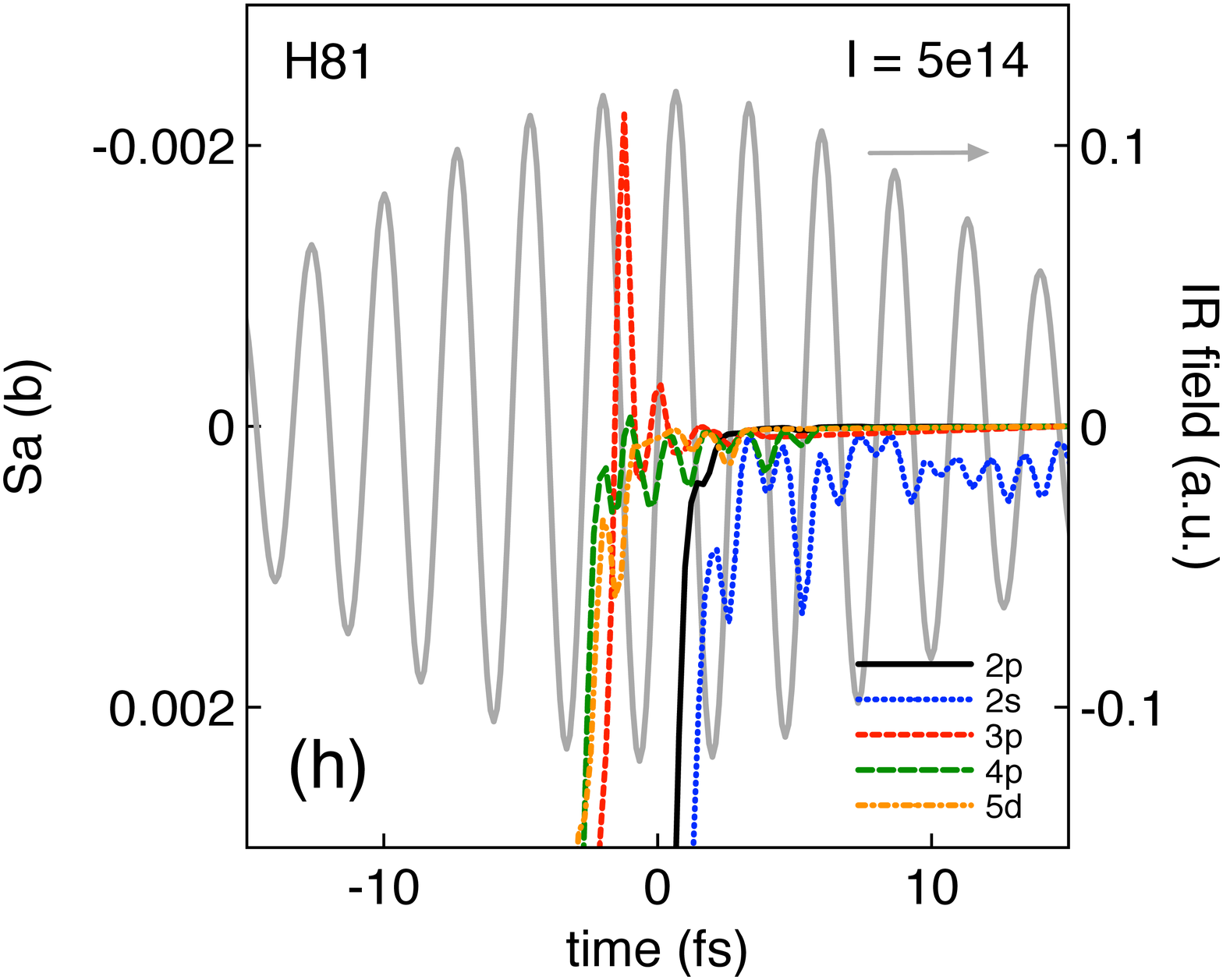}  
      \includegraphics[width=0.32\textwidth]{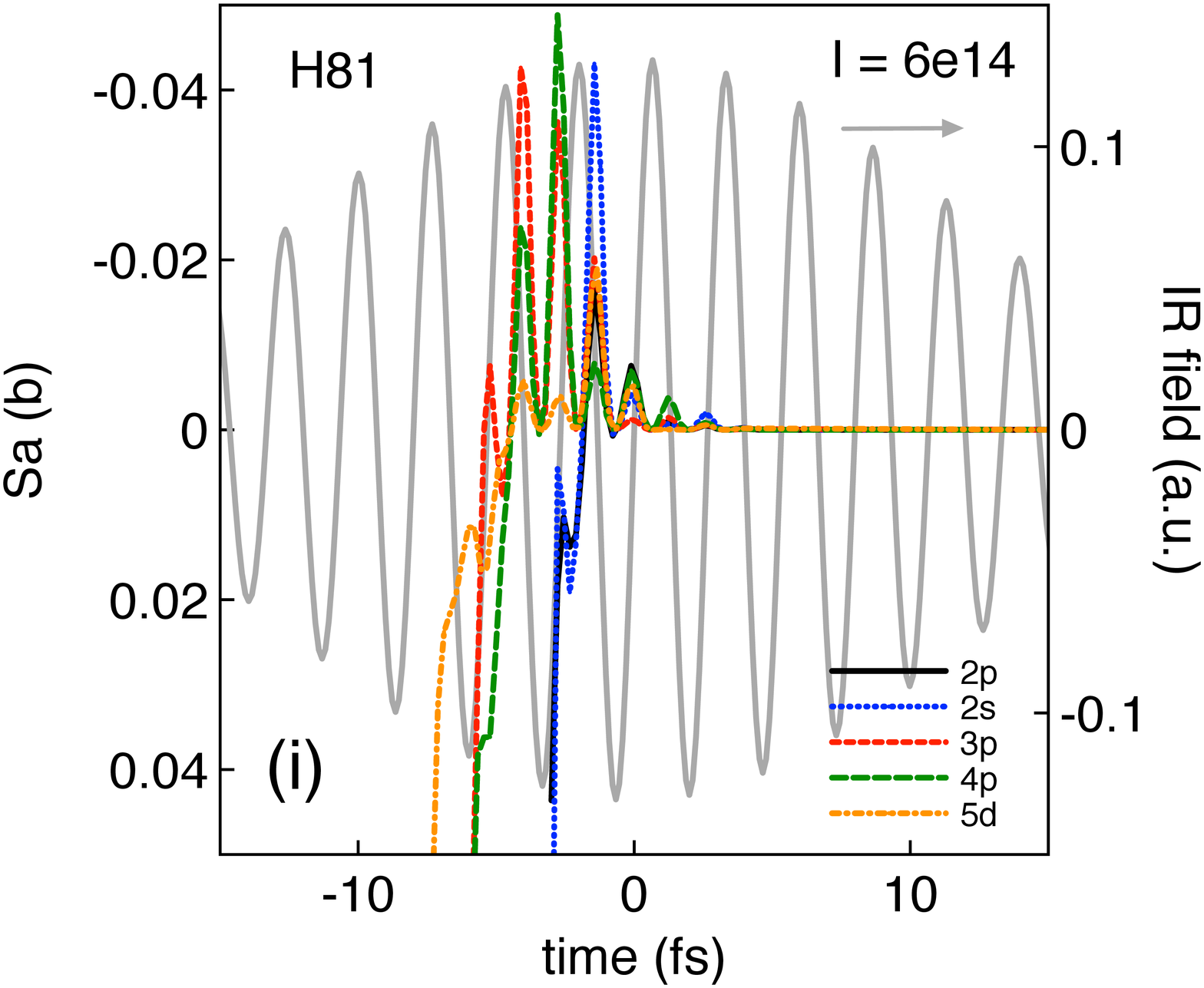}
\end{minipage}\qquad
\caption{Spectrally integrated single-atom XUV absorption signal $S_a$ (in barns) as a function of the delay (time) between the 
XUV and the IR pulses. The calculations are performed for H61, H71 and H81, and the IR peak intensities 
$4\times10^{14}$ W/cm$^2$, $5\times10^{14}$ W/cm$^2$, and $6\times10^{14}$ W/cm$^2$, as indicated. The
different lines in each plot correspond to the system initially in the He$^+$ ion 2p, 2s, 3p, 4p, and 5d bound states. The 
vertical left axis is reversed for clarity.}
\label{fig2}
\end{figure*}

\section{Numerical simulations}

The main results of our simulations are shown in Fig. \ref{fig2}. We have considered a linearly polarized 
IR field resulting from a 20-cycle cos$^2$ envelope vector potential ($\approx$20 fs duration) of 800 nm, with IR field peak 
intensities of 4.0 -- 6.0  $\times$10$^{14}$ W/cm$^2$, as indicated in the subplots in Fig. \ref{fig2}. 
The IR field interacts with an He$^+$ ion together with a weak ($\sim $10$^{12}$ W/cm$^2$) 
XUV single pulse of cos$^2$ shape and of $\approx$ 0.55 fs intensity FWHM duration, 
considering the harmonic frequencies H61, H71 and H81, as it is also shown in the figure.
We have considered first the study involving a single attosecond XUV pulse in order to conveniently 
analyze the physics of the amplification process. We later compare these results to the measurements of the experiments and
to calculations using a train of pulses.
In Fig. \ref{fig2}, $t=0$ is considered at the center of the sin--IR pulse, so that the delay between the XUV and the IR pulse 
is given by the central time position of the XUV pulse and coincides to the values in the horizontal time-axis. 

In order to analyze the absorption/gain of the XUV pulse in the medium we make use of a regularly used technique,
that is we compute the total probe absorption signal $S_a$ obtained in the frame of transient absorption as described 
in \cite{Mukamel1} (see e.g. \cite{Serrat6} for further details).
In all calculations atomic units (a.u.) are considered and the total probe absorption signal $S_a$ is 
given in units of area (barns). Positive values of $S_a$ mean absorption of the XUV pulse in the single-atom medium,
while negative values mean amplification (gain), so that the absorption cross section $S_a$ 
multiplied by the number density of atoms results in the absorption/gain coefficient \cite{Mukamel1}.
Our study is centered in the 90 -- 130 eV XUV gain region that has been 
observed in recent HHG measurements in He. 
As commented above, our simulations indicate that if the population density of excited states is significantly low 
compared with the ground state He$^+$ (1s) ion density, only absorption is obtained at all delays. Thereupon, 
we have considered the He$^+$ ion initially in diverse excited states such as 2p, 2s, 3p, 4p and 5d. 
Convergence of the numerical simulations has been verified in all cases.

Having a look at Fig. \ref{fig2}, we observe that several gain regions are produced at different delays
for the central photon energies H61, H71 and H81. Those gain regions for such high photon energies 
had not been predicted before by using the solution of a first principles theory at the single-atom level. 
The degree of the amplification depends both on the central XUV photon energies and the peak intensity of the
IR pulse considered. In the simulations, when the He$^+$ ion is initially in one of the 2p, 2s, 3p, 4p or 5d bound states,
it first interacts with the leading part of the IR pulse, which can tunnel ionize the He$^+$ ion
and single and multi-photon excitations to the lower and upper bound states can also occur. 
Quite surprisingly, although the duration of the IR pulse is in principle enough to allow amplification in several IR cycles,
the synchronized XUV seed pulse is only amplified in a particular cycle -- or at most in two
consecutive cycles, of the IR pulse. As it will be shown below, this fact is in agreement with the experimental 
measurements, and it shows that only when the excited states of the He$^+$ ion are optimally depleted by the 
IR pulse, stimulated recombination from continuum states by the synchronized XUV seed pulse occurs.  
Otherwise, the details of the regions where the amplification is produced in 
Fig. \ref{fig2} are difficult to disentangle. We observe the highest gain for H61 of about 30 b at the higher 
IR peak intensity that we have considered (6$\times$10$^{14}$ W/cm$^2$) [see Fig. \ref{fig2} (c)], which occurs 
for the He$^+$ ion being initially in the 3p state and at the earlier delay of all cases ($\approx -9.5$ fs). This can be
expected since for the higher IR peak intensities the excited states of the He$^+$ ion are depleted earlier.
We also observe that the gain decreases as the central XUV pulse photon energy is increased, and it is higher for the 
higher IR peak intensities, which allows us to predict a gain of $\sim$0.05 b with 6$\times$10$^{14}$ W/cm$^2$ 
at photon energies as large as H81 ($\approx 125.5$ eV). Our results hence show how the 
amplification of a particular photon energy region can be precisely tuned 
by varying the peak intensity of the IR laser pulse. 
It is also worth noting that the peak intensities typically used in several
experiments ($\sim 10^{15}$ W/cm$^2$) can be higher than the ones that we have considered, and that 
the IR pulse durations can also typically be longer. Other numerical studies considering higher peak 
IR intensities with 20-cycle or longer durations are however almost prohibitive due to computational costs.

From our simulations it can therefore be inferred why having He$^+$ ions in excited states results 
in a decisive effect for XPA processes at energies about 100 eV to be efficient using He
as amplifying medium. The electrons that recombine by HHG to the ground
state of He$^+$ and produce high order harmonics around 100 eV need to carry approximately the 
I$_p$ He$^+$ energy (54.4 eV). In other words, the states in the continuum from which the recombination 
is produced need to be of an energy of this order. In the case of recombination to the ground state of He, however, the
energy in the continuum needs to be much higher to reach 100 eV, since the I$_p$ for He is only 24.6 eV, 
and that clearly reduces the probability for parametric stimulated recombination. 
A classical trajectory analysis shows that other 
parametric channels such as impulsive parametric IXPA as described in \cite{Serrat3,Serrat4,Serrat6} might 
also be open for amplification at the high photon energy regions that we investigate, although these processes 
would only be effective for transitions from the ground state of He$^+$, which makes them almost negligible 
in the present IR configuration. 

\begin{figure*}[ht]
\begin{center}
\includegraphics[scale=0.42,clip=true,angle=0]{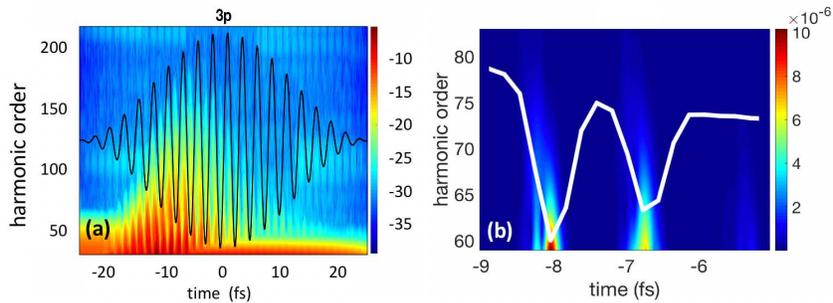}
\end{center}
\caption{(a) Frequency-time analysis of the high-order harmonics intensities generated by an IR alone 
with peak intensity of 5$\times$10$^{14}$ W/cm$^2$, in the case that the system is initially in the 3p bound state. The
intensity of the generated harmonics is shown in logarithmic scale and arbitrary units. 
(b) Zoom of the frequency-time analysis in (a) close to the harmonic H61. The
intensity of the generated harmonics is here in linear scale and arbitrary units. 
The calculated spectrally integrated single-atom XUV absorption signal $S_a$ for H61 
(Fig. \ref{fig2} (b) -- red-dashed line) is plotted on top of the spectra (white line). The
vertical axis of this white line has been reversed here with respect to Fig. \ref{fig2} (b) for a better visual comparison.}
\label{fig3}
\end{figure*}

We next explicitly analyze the amplification process which is unambiguously
described as a stimulated recombination process from the continuum states.
Figure \ref{fig3} (a) shows a frequency-time analysis of the high-order harmonics generated by an IR pulse alone of  
peak intensity 5$\times$10$^{14}$ W/cm$^2$, in the case that the system is initially in the 3p state, with the IR electric field
plotted on top for clarity (black line). We observe how the main high order harmonics are produced at the leading part or the IR pulse.
Figure \ref{fig3} (b) is an enlargement of Fig. \ref{fig3} (a) near the H61 harmonic in the delay region between $-$9 and $-$5 fs,
and the calculated absorption signal $S_a$  (Fig. \ref{fig2} (b) -- red-dashed line) is plotted on top of the spectra (white line).
Definitely, the gain of the H61 XUV pulse is formed precisely at the same times ($\approx -$8.1 fs and $\approx -$6.8 fs) 
as the H61 harmonics from HHG are produced, which is the first theoretical demonstration of parametric XPA processes
using accurate first principles simulations. This allows us to corroborate that XPA
is the physics leading to the observed XUV amplification at photon energies far from the ionization 
threshold observed in He gas \cite{Seres1,Seres2,Seres3,Seres4,Serrat3}. We have also checked that this is 
essentially the case for all the gain peaks shown in Fig. \ref{fig2}.
\begin{figure*}[ht]
\begin{center}
\includegraphics[scale=0.33,clip=true,angle=0]{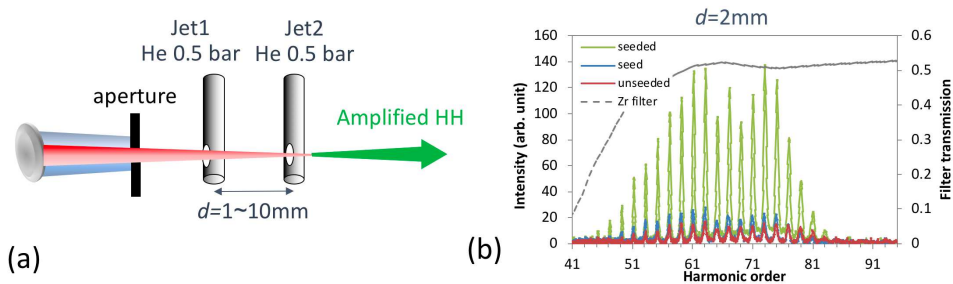}
\end{center}
\caption{(a) Experimental setup of the two-jet arrangement. (b) Harmonic spectra measured from the jets separately 
(seed and unseeded) and the two jets together (seeded) with the transmission curve of the used Zr thin-film filter.}
\label{fig4}
\end{figure*}

\section{Comparison with the experimental measurements}

In the experiments, in order to measure the delay dependent gain in He, a Ti:sapphire laser system was used. 
80 fs pulses (central wavelength of 800 nm) were focused by a spherical mirror with focal length of 2500 mm to a double-jet arrangement, as shown in 
Fig. \ref{fig4}(a). In the focus, the beam waist was $\approx$250 $\mu$m and the laser intensity was 
$\approx$5$\times$10$^{14}$ W/cm$^2$. The first He gas jet (Jet1) produced the seed pulse for the second amplifier He jet (Jet2). Typical spectra obtained by the different jet combinations can be seen in Fig. \ref{fig4}(b) for Jet1 alone (seed), for Jet2 alone (unseeded) and the two jets together (seeded). 
The spectrum of the two jets together is much more intense than the seed 
and unseeded spectra and cannot be obtained by a coherent superposition. 
\begin{figure*}[ht]
\begin{center}
\includegraphics[scale=0.45,clip=true,angle=0]{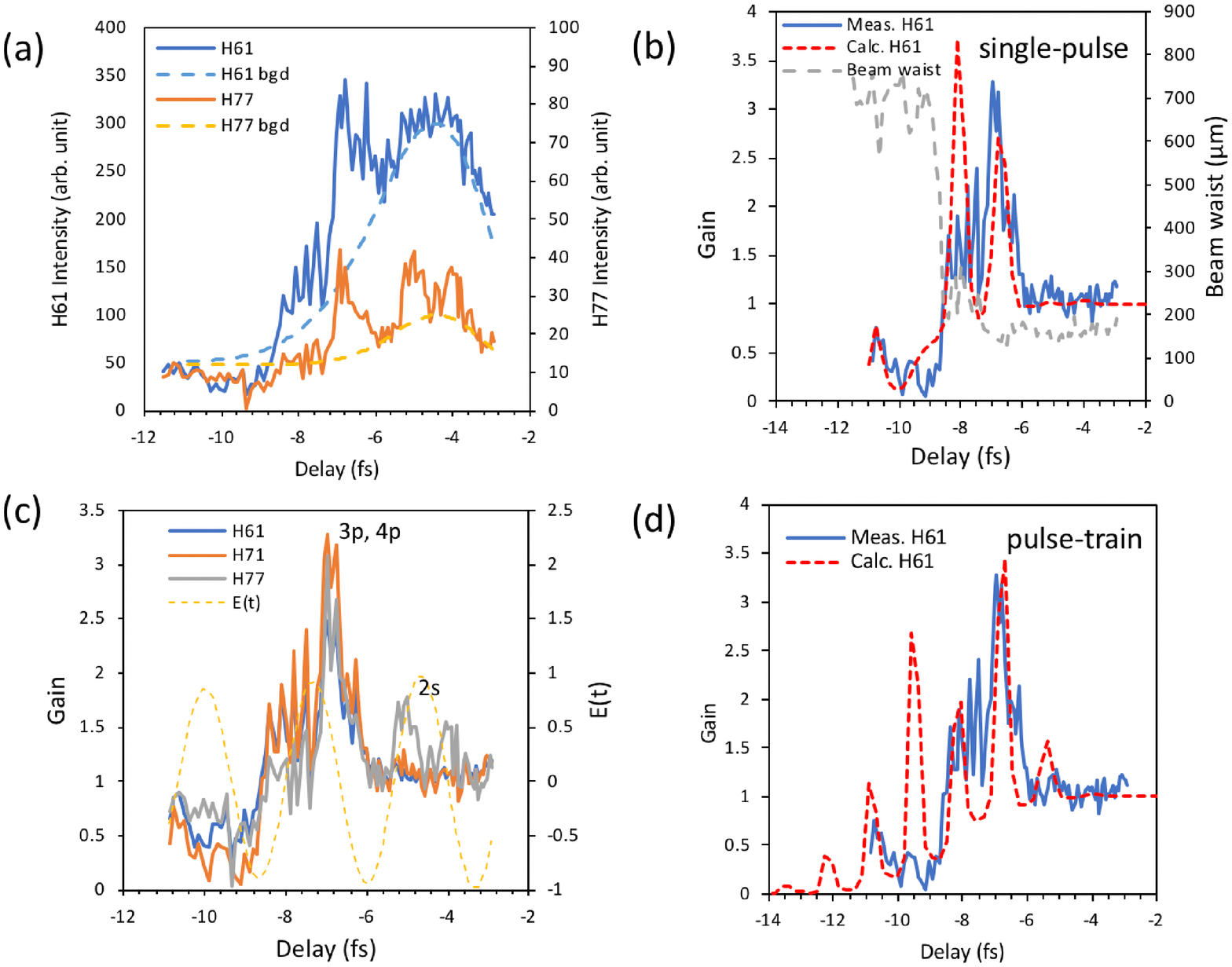}
\end{center}
\caption{(a) Delay dependence of two representative harmonic lines, H61 and H77, with the fitted reference Gaussian shapes. 
(b) Comparison of the measured and the calculated parametric gain together with the harmonic beam waist. 
As it is clear, the gain occurs at the same time as the beam waist decreases. 
(c) Measured gain dynamics at three harmonic lines: H61, H71 and H77.
(d) Comparison of the measured and the calculated parametric gain considering a train of H61 pulses as seed.}
\label{fig5}
\end{figure*}
To reveal the gain dynamics in He gas, the delay between the laser pulse and the HHG seed (or probe) pulse can be scanned by changing the distance 
between the two jets so that the caused delay range has to be comparable to one optical IR cycle.
The delay has two possible origins \cite{Seres1,Seres4}, one of them is the difference of the Gouy phases at the positions of the gas jets within the focused laser beam and the other is the effect of the free electrons to the beam propagation between the two jets.
Figure \ref{fig5} shows the main dynamics observed from the measurements and the comparison with the simulations.
In the measurements, the scanning was made in 20 $\mu$m steps within a range of 2 mm. Because this range is much smaller than the $\sim$50--60 mm Rayleigh length of the laser beam used, the contribution of the Gouy phase represents a negligible delay ($<$ 30 as).
A much larger effect is given by the contribution of the free electrons to the refractive index during the propagation of the 
laser beam between the jets. The delay $\tau$ given by the distance is $\tau \approx (e^2 d_0 n_e [1-{\rm exp}(-d/d_0)])/(2\epsilon_0 m_e c\omega^2 )$,
where $n_e = n_{e1} + n_{e2}$ is the sum of the free electron densities of the contributing two gas jets, 
$d$ is the distance between the jets, $d_0$ is the distance from the center of the gas jets to where their gas pressure drops to 1/e, 
and $\omega$ is the laser frequency, with the rest of parameters being the common SI constants. 
The delay is scaled in Fig. \ref{fig5} as in the calculations in Fig. \ref{fig2} by using the value $d_0 = 2.3$ mm 
and 28\% free electron density at 0.5 bar gas pressure.
A Gaussian shape has been fitted in Fig. \ref{fig5}(a) (dashed lines) to serve as a reference 
for the additional propagation effects, which are not considered in the theory.
Intensity peaks can then be clearly identified as the indication of the appearance of gain at very specific XUV--IR delays.

Beyond the spectral shape, the beam profiles of the harmonic beams were also evaluated from CCD images of the spectra. 
The gray dashed line in Fig. \ref{fig5}(b) shows the beam waist for the most intense harmonic order that was measured (H61). 
It is clear that the XPA feature appears together with an almost abruptly decrease of the beam waist at about $-$9 fs delay.  
Dividing the measured pulse by the reference Gaussian shape, the observed gain for H61 is plotted in Fig. \ref{fig5}(b). 
The measured and the calculated gains have been scaled and are plotted together for comparison.  
The strongest peak at about $-$7 fs delay and the somewhat weaker one at about $-$8 fs delay can be attributed to the 3p 
orbitals of the excited He$^+$ ions by comparing the results with the calculations in Fig. \ref{fig2}(b). 
As it can be expected, the two gain peaks are separated by half an IR cycle, and remarkably both theory and measurements 
show parametric amplification only at a few specific cycles of the IR laser pulse. Since our simulations do not consider 
propagation macroscopic effects, only a qualitative agreement can be expected between the simulations and the measurements. 

Figure \ref{fig5}(c) shows together the measured gain dynamics at the harmonic lines H61, H71 and H77. In the measurements the main peaks appear at 
the same delays for the three different harmonics, which differs from the simulations as they show up there at consecutive cycles [compare e.g.
Fig. \ref{fig2}(b) with Fig. \ref{fig2}(e)]. This discrepancy can be attributed to the longer duration of the IR pulse used in the experiments (80 fs) compared with 
the shorter pulse (20 fs) used in the simulations, since the 80 fs pulse includes consecutive IR cycles with much more similar strength. As a matter of course,
the theory hence shows that shorter IR pulses are preferred for an accurate tuning of the amplified harmonics region in the present configuration.
The peaks at about $-$5 fs and $-$4 fs delays in Fig. \ref{fig5}(c) for H77 can be observed in Fig. \ref{fig2}(e)
for H71, and therefore they might be attributed to generation of harmonics from the 2s initial state of He$^+$, as noted in the figure. 
Finally, Fig. \ref{fig5}(d) is a comparison of the measured and the calculated gain dynamics considering a seed train of H61 pulses separated 
by half IR-cycle, which is closer to the interaction that is produced in the experiments.  
Essentially, the phenomena that we have described with single XUV pulse remain, although some more IR cycles are in this case involved
in the amplification. We believe that this additional peaks might be caused by the interplay of the several pulses 
of the train in the interaction with the He$^+$ ion, which changes the depletion of the corresponding excited states. As commented above, however, 
a more accurate comparison between theory and measurements necessarily needs to include macroscopic propagation effects, which requires exceptional computational costs.  

\section{Conclusions}

In conclusion we have shown that amplification of high photon energy XUV pulses far from the 
ionization threshold (around 100 eV) can be described at the single-atom level by the 3D time-dependent 
Schr{\"o}dinger equation in He$^+$, which allows us to unambiguously reveal both in the time and the frequency domains 
how the physics behind these amplification processes have a parametric character.  
In accordance with the experiments, we corroborate that the XUV pulses need to be perfectly synchronized in 
time with the driving laser field to produce stimulated recombination. 
Remarkably, the gain appears only within a narrow time window, mainly within one optical cycle delay.
We have demonstrated that He$^+$ has to be in some excited state to produce gain
and that transitions from the continuum to the ground state of He$^+$ ions are essential for efficient photon energy amplification far from the 
ionization threshold. Our study reveals how the gain and photon energy for XPA can be controlled by the intense IR pulse parameters
and indicates the basic physics for future experiments, so that the desired attosecond XUV gain needed for applications, such as intense 
coherent seeds for plasma x-ray lasers, can be achieved.

\section*{Funding}
Spanish Ministry of Economy and Competitiveness through FIS2017--85526--R;
Japan Society for the Promotion of Science (JSPS) (JP17H02813).

\section*{Acknowledgments}
Computation provided by the Super Computer System, Institute for Chemical Research, Kyoto University, is acknowledged.

\end{document}